\input ./style/arxiv-general.cfg
\documentclass[aoas,MSNbibl,nameyear,rotating,dvips]{arximspdf}
\makeatletter
   \@ifpackageloaded{graphicx}{}{\usepackage{graphicx}}
\makeatother
\usepackage{dcolumn}

\doi{10.1214/15-AOAS874}
\volume{10}
\issue{1}
\pubyear{2016}
\firstpage{94}
\lastpage{117}
\docsubty{FLA}

\makeatletter
\newcolumntype{d}[1]{D{.}{.}{#1}}
\newproclaim{proc}{Procedure}
\makeatother

\begin{document}
\begin{frontmatter}

\title{Prevalence and trend estimation from observational data with
highly variable post-stratification weights\thanksref{T11}}
\runtitle{Prevalence and trend estimation from observational data}

\begin{aug}
\author[A]{\fnms{Yannick}~\snm{Vandendijck}\corref{}\thanksref{M1,T1}\ead[label=e1]{yannick.vandendijck@uhasselt.be}},
\author[A]{\fnms{Christel}~\snm{Faes}\thanksref{M1,T2}\ead[label=e2]{christel.faes@uhasselt.be}}
\and
\author[B]{\fnms{Niel}~\snm{Hens}\thanksref{M1,M2,T3}\ead[label=e3]{niel.hens@uhasselt.be}}
\runauthor{Y. Vandendijck, C. Faes and N. Hens}
\affiliation{Hasselt University\thanksmark{M1} and University of
Antwerp \thanksmark{M2}}
\address[A]{Y. Vandendijck\\
C. Faes \\
Interuniversity Institute for Biostatistics \\
\quad and Statistical Bioinformatics \\
Hasselt University \\
3590 Diepenbeek\\
Belgium\\
\printead{e1}\\
\phantom{E-mail: }\printead*{e2}}
\address[B]{N. Hens \\
Interuniversity Institute for Biostatistics \\
\quad and Statistical Bioinformatics \\
Hasselt University \\
3590 Diepenbeek\\
Belgium\\
and\\
Centre for Health Economic Research\\
\quad and Modeling Infectious Diseases \\
Vaccine and Infectious Disease Institute \\
University of Antwerp \\
2610 Wilrijk\\
Belgium\\
\printead{e3}}
\end{aug}
\thankstext{T11}{Supported in part by the IAP Research Network P7/06 of
the Belgian State (Belgian Science Policy).}
\thankstext{T1}{Supported in part by a doctoral grant of Hasselt
University, BOF11D04FAEC.}
\thankstext{T2}{Supported in part by the National Institutes of Health
award number R01CA172805.}
\thankstext{T3}{Supported in part by the University of Antwerp
scientific chair in Evidence-based Vaccinology,
financed in 2009--2014 by a gift from Pfizer.}

%
\received{\smonth{12} \syear{2014}}
%
\revised{\smonth{7} \syear{2015}}

%
\begin{abstract}
In observational surveys, post-stratification is used to reduce bias
resulting from differences
between the survey population and the population under investigation.
However, this can lead to inflated post-stratification weights and,
therefore, appropriate methods are required to obtain less variable
estimates. Proposed methods include collapsing post-strata, trimming
post-stratification weights, generalized regression estimators (GREG)
and weight smoothing models, the latter defined by random-effects
models that induce shrinkage across post-stratum means. Here, we first
describe the weight-smoothing model for prevalence estimation from
binary survey outcomes in observational surveys. Second, we propose an
extension of this method for trend estimation. And, third, a method is
provided such that the GREG can be used for prevalence and trend
estimation for observational surveys. Variance estimates of all methods
are described. A simulation study is performed to compare the proposed
methods with other established methods. The performance of the
nonparametric GREG is consistent over all simulation conditions and
therefore serves as a valuable solution for prevalence and trend
estimation from observational surveys. The method is applied to the
estimation of the prevalence and incidence trend of influenza-like
illness using the 2010/2011 Great Influenza Survey in Flanders, Belgium.
\end{abstract}

%
\end{frontmatter}

\section{Introduction}
\label{sec1}
Frequently, researchers are interested in estimating the overall mean
or time trend of an outcome in the study population based on a survey.
To obtain valid estimates, the sample must be representative for the
population. Stratified sampling provides samples that are
representative for the population and can reduce the variance of the
estimators. In observational surveys, though, it is not possible to
perform stratified sampling. However, when auxiliary information is
available, post-stratification can be used to correct for known
differences between the obtained sample and the population under
investigation. This is done by equating the sample distribution of a
secondary variable with its distribution in the population, and
adjusting estimates using appropriate weighting techniques. When the
secondary variable is related with the survey outcome variable,
post-stratification can improve both the accuracy and precision of
estimators [\citet{Little1993}]. In this article we restrict
ourselves to the setting where a binary survey outcome is of interest
and an ordinal (or interval-scaled) post-stratifying variable is available.

Post-stratification weights play important, but different roles in
design-based and model-based inference. In design-based inference, it
is assumed that the sample inclusion indicator is random and the survey
variables are fixed. Popular design-based estimators are the
Horvitz--Thompson (HT) estimator and its extensions which weigh
observations by the inverse of their probability of inclusion
[\citet{Horvitz1952}]. Although these estimators are design
consistent [\citet{Isaki1982}], they can be very inefficient when
highly variable post-stratification weights are present, as illustrated
by Basu's (1971) famous elephant example. \nocite{Basu1971}

In contrast, in model-based inference, distributional assumptions are
made about the survey outcome and a model is used to predict the
outcome for the nonsampled units [\citet{Little2004}]. Such
model-based prediction estimators are consistent, but are subject to
bias when the underlying model is misspecified. Weight smoothing models
treat the strata of the post-stratifying variable as random effects in
the model. In this manner, information between strata is borrowed
through shrinkage related to the sample size in each stratum and,
consequently, post-stratification weights are implicitly smoothed
[\citet{Elliott2000}]. In the literature, these models are
extensively described for Gaussian data within the empirical Bayesian
or full Bayesian framework
[\citet{Ghosh1986,Lazzeroni1998,Elliott2000,Gelman2007},
\citeauthor{Little1983} (\citeyear{Little1983,Little1991,Little1993}), \citet{Zheng2004}].
In the full Bayesian framework, generalized linear regression
estimators for non-Gaussian data have been discussed [\citet{Elliott2007}].

Weight smoothing models are similar to model-based prediction models
widely used in the field of small area estimation (SAE) [\citet
{Rao2003}]. In SAE, generalized linear mixed models are used for the
prediction of small area proportions, with random effects introduced to
capture area specific differences. Because of the similarity between
these models and weight smoothing models, several results from SAE are
also applicable in the context of weight smoothing models.

\citet{Lehtonen1998} proposed the generalized regression estimator
(GREG), which is a design-based model-assisted estimator where
predictions based on a suitable model and the HT estimator for the
model residuals are combined. This GREG estimator was extended by
\citet{Chen2010} who introduce a Bayesian penalized spline
predictive estimator for probability-proportional-to-size sampling.
Their predictive model includes the weights as covariates.

In this paper, focus is on estimating the prevalence of a survey
outcome when one or more post-stratification weights are large. A
weight smoothing model based on a generalized linear mixed model is
used for this purpose. This weight smoothing model for binary data is a
natural extension of the weight smoothing model for continuous data.
The predicted values from the model are used in the GREG estimator. The
two major additions of this article to the existing literature are as
follows: ({i}) the adjustment of the weights used in the GREG estimator
such that the GREG estimator can be used for post-stratification
weights obtained from an observational survey; and ({ii}) a weight
smoothing model for estimating a time trend of a binary survey outcome.

The outline of this paper is as follows. Section \ref{sec2} describes a
motivating data example that concerns the estimation of the overall
prevalence and weekly incidence trend of influenza-like illness from
the Great Influenza Survey (GIS) in Flanders, Belgium. Notation and
conventional method of analysis are given in Section \ref{sec3}.
Section \ref{sec4} describes our modeling approaches, and in Section
\ref{sec5} these methods are compared with standard alternatives in an
elaborate simulation study. We return to the GIS data example in
Section \ref{sec6}, and conclude the paper with a discussion in Section
\ref{sec7}.

\section{Motivating example}
\label{sec2}
The Great Influenza Survey (GIS) is an observational survey based on
the voluntary participation of individuals via the internet aiming at
monitoring influenza-like illness (ILI) in the Netherlands and Belgium
[\citet{Marquet2006,Friesema2009,Vandendijck2013}]. Participants
of the GIS receive a weekly email with a link to a questionnaire (it is
not mandatory to complete this questionnaire). Based on the
questionnaire, well-defined criteria are used to determine whether or
not the participant has experienced an ILI episode in the preceding
week. At the start of the flu-season (or at registration), an
additional questionnaire needs to be completed to obtain information on
several demographic characteristics. In this paper, we focus on the GIS
in Flanders, Belgium, during the 2010--2011 influenza season. Interest
is in the estimation of the overall mean prevalence and the weekly
incidence trend of ILI. Data from the Flemish GIS can be obtained upon
request via \surl{www.influenzanet.com}.

The GIS recorded information of 27 weeks during the 2010--2011 influenza
season (from 1/11/2010 up to 8/5/2011). In total, there were 4551
participants, with a total of 83,500 records. The highest number of
respondents in a week was 3467 and the minimum was 2402.

\begin{figure}

\includegraphics{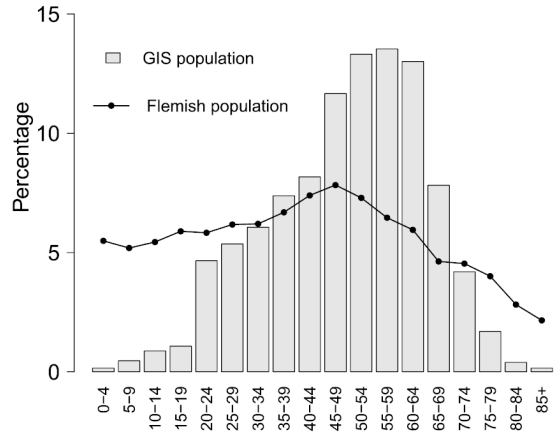}

\caption{Age distribution of the Great Influenza Survey (GIS)
population and the overall Flemish population stratified by 18 age
intervals of five years.}
\label{Figure1}
\end{figure}

The age of the participants ranged from 1 to 88 years. From Figure \ref
{Figure1} it is observed that the age distribution of the survey
population is very dissimilar to the age distribution of the overall
Flemish population and, therefore, we post-stratify the survey
according to 18 five-year age intervals. The population age
distribution of the overall Flemish population is obtained from
Statistics Belgium (\surl{www.statbel.fgov.be}). Some post-strata are seriously
underrepresented and therefore high post-stratification weights are
present, with weights ranging from 0.46 up to 35.70 (in Section \ref
{sec3} a definition of post-stratification weights is presented).

In Section \ref{sec6}, results for this survey are given to illustrate
the methods described in this paper. In this analysis, we will assume
no response bias is present. Because this data example is only used for
illustration purposes, we do not correct for gender
post-stratification, though the method allows easy extension to take
this into account.

\section{Notation and the conventional methods of analysis}
\label{sec3}
\subsection{Notation}
\label{sec31}
Let $Y$ denote a binary survey outcome variable and $X$ an ordinal
post-stratifying variable with $H$ levels and known population
distribution. Let $N_{h}$ denote the population size and $n_{h}$ the
sample size in post-stratum~$h$, for $h=1,\ldots,H$. We assume that
$N_{h}$ is obtained from auxiliary data and $n_{h}>0$ ($\forall
h=1,\ldots,H$). The total population and sample size are denoted by
$N=\sum_{h=1}^{H}N_{h}$ and $n=\sum_{h=1}^{H}n_{h}$, respectively. It
is assumed that the respondents in a post-stratum can be treated as a
random sample. Interest is in the estimation of (or inference about)
the population prevalence, namely,
$
\bar{Y}=\frac{1}{N}\sum_{i=1}^{N}Y_{i}=\sum_{h=1}^{H}P_{h}\bar{Y}_{h},
$
where $\bar{Y}_{h}$ is the population mean and $P_{h}=N_{h}/N$ is the
population proportion of post-stratum $h$.

\subsection{Design-based methods}
\label{sec32}
Standard design-based approaches to estimate $\bar{Y}$ consider
estimates of the form
$
\bar{y}=\frac{1}{n}\sum_{i=1}^{n}w_{i(h)}y_{i},
$
where $w_{i(h)}$ is the weight of observation $i$ belonging to
post-stratum $h$ with $\sum_{i=1}^{n}w_{i(h)}=n$, and $y_{i}$ is the
corresponding observed survey outcome. Some special cases are discussed.

\begin{longlist}[(2)]
\item[(1)] The \emph{unweighted sample prevalence}, $\bar{y}_{\mathit{unw}}$, is
obtained when $w_{i(h)}=1$ ($\forall i=1,\ldots,n$). It can be written
as $\bar{y}_{\mathit{unw}}=\sum_{h=1}^{H}p_{h}\bar{y}_{h}$, with $p_{h}=n_{h}/n$
the sample proportion and $\bar{y}_{h}$ the sample prevalence in
post-stratum $h$. It is an unbiased estimate whenever $Y$ and $X$ are
unrelated or as long as the probability of inclusion does not depend on
$X$. However, if sampling variability is apparent or systematic bias in
the sampling procedure is present, $p_{h}$ deviates from its population
counterpart $P_{h}$ and the unweighted sample prevalence is a biased
estimator for $\bar{Y}$.

\item[(2)] The \emph{post-stratified prevalence}, $\bar{y}_{\mathit{ps}}$, is obtained
when $w_{i(h)}=P_{h}/p_{h}$ ($\forall i=1,\ldots,n$) and can be written
as $\bar{y}_{\mathit{ps}}=\sum_{h=1}^{H}P_{h}\bar{y}_{h}$. The $w_{i(h)}$ are
called post-stratification weights in this case. This is an unbiased
estimate of $\bar{Y}$. However, the post-stratified prevalence can be
unstable because the sample prevalence in post-stratum $h$ is used in
$\bar{y}_{\mathit{ps}}$ regardless of the number of respondents in post-stratum~$h$. Second, when a post-stratum contains only few observations, the
variance is inflated. This inflated variance can overwhelm the bias
reduction, so that the post-stratified prevalence yields an increased
mean-squared error.

\item[(3)] In order to solve the instability and inflated variance of $\bar
{y}_{\mathit{ps}}$, a \emph{weight-trimming prevalence} estimator, $\bar
{y}_{\mathit{trim}} $, obtained by trimming all weights larger than a
prespecified cutoff value $w_0$, can be used. The other weights are
adjusted to maintain the same weighted sample size. The trimmed
prevalence estimate can be written as
$
\bar{y}_{\mathit{trim}} = \sum_{h=1}^{l-1}\gamma\frac{N_{h}}{N}\bar{y}_{h} +
\sum_{h=l}^{H}w_{0}\frac{n_h}{n}\bar{y}_{h},
$
with normalizing constant
$
\gamma= \frac{n-\sum_{i}\zeta_{i}w_0}{\sum_{i}(1-\zeta_{i})w_{i(h)}},
$
where $\zeta_{i}$ equals 1 when $w_{i(h)}>w_0$ and 0 otherwise. The
choice of the cutoff value $w_{0}$ is mostly done \textit{ad hoc},
although \citet{Potter1990} presented several systematic methods
for choosing $w_{0}$ based on the data.

Other methods to solve the instability and inflated variance of $\bar
{y}_{\mathit{ps}}$ include pooling or collapsing small post-strata that
contribute excessively to the variance [\citet
{Tremblay1986,Little1993}]. Weight trimming can be seen as a special
case of collapsing several post-strata. \citet{Elliott2000}
proposed an extension of the method, the compound weight pooling
method, by conducting the pooling at every possible level of $w_{0}$
and computing a Bayesian weighted average.
\end{longlist}

\section{Methodology}
\label{sec4}
\subsection{Model-based inference for prevalence estimation}
\label{sec41}
Design-based methods ignore the ordinal nature of the post-stratifying
variable, whereas a model-based approach reflects the intrinsic order
and allows to borrow\break strength from neighboring strata with more
information. Weight smoothing models directly model the means in the
weight strata making use of random effects [\citet
{Lazzeroni1998}]. The weight smoothing model for a binary survey
outcome is
%
\begin{equation}
\label{eqgeneralmodel} y_{i(h)}|\mu_{h} \sim\operatorname{Bern}(
\mu_{h})\quad\mbox{and}\quad \bolds{\delta}^{*} \sim
\mathcal{N}_{H}(\bolds{\delta},\mathbf{D}),
\end{equation}
where $g(\mu_h) = \delta^*_h$, $\bolds{\delta}^{*} = (\delta
_{1}^{*},\ldots,\delta_{H}^{*})^{T}$ and $\bolds{\delta}=(\delta
_{1},\ldots,\delta_{H})^{T}$ are vectors of unknown parameters, $\mathbf
{D}$ is an unknown $H \times H$ variance--covariance matrix and $g(\cdot)$
is the logit-link function. The following choices for $\bolds{\delta}$
and $\mathbf{D}$ have been considered in model (\ref{eqgeneralmodel}):

\begin{longlist}[(a)]
\item[(a) \textit{Exchangeable random effects} (\textit{XRE}):] $\delta_{h} = \beta$
for all $h$ and $\mathbf{D} = \sigma^2\mathbf{I}_{H}$
[\citet{Holt1979,Ghosh1986}, \citeauthor{Little1983} (\citeyear{Little1983,Little1991})].

\item[(b) \textit{Linear model} (\textit{LIN}):] $\delta_{h} = \beta_{0}+\beta
_{1}X_{h}$ for all $h$ and $\mathbf{D} = \sigma^2\mathbf{I}_{H}$
[\citet{Lazzeroni1998}].

\item[(c) \textit{Nonparametric} (\textit{NPAR})] $\delta_{h} = f(X_{h})$ for all $h$
and $\mathbf{D} = \sigma^2\mathbf{I}_{H}$, where $f$ is a nonparametric
spline function [\citet{Elliott2000,Zheng2004}]. We use the
approximating thin plate spline family for constructing $f$. This
choice was made because this spline family is readily available in the
procedure \mbox{\textit{GLIMMIX}} in the software package SAS. Simulations
indicate that the approximating thin plate splines perform as well as
linear truncated splines or cubic B-splines (see Supplementary Material [\citet{supp}]).
\end{longlist}

All the above models can be cast in the generalized linear mixed model
(GLMM) framework:
$
g(E(\mathbf{y}|\mathbf{b})) = \bolds{\eta} \equiv\mathbf{N}\mathbf
{X}\bolds{\beta} + \mathbf{N}\mathbf{Z}\mathbf{b},
$
where $\mathbf{N}$ is an $n \times H$ matrix indicating to which
stratum an observation belongs [$(\mathbf{N})_{ih}=1$ if $y_{i}$
belongs to stratum $h$ and $0$ otherwise], $\mathbf{X}$ is an $H \times
p$ fixed-effect design matrix and $\bolds{\beta}$ is a $p \times1$
vector of unknown fixed-effect parameters, $\mathbf{Z}$ is an $H \times
q$ random-effect design matrix and $\mathbf{b}$ is a $q \times1$
vector of unknown random effects for which it is assumed that $\mathbf
{b} \sim\mathcal{N}_{q}(\mathbf{0},\mathbf{G})$. In Appendix~A of the
Supplementary Material [\citet{supp}], it is shown how models (a)--(c) are formulated
in the GLMM framework. Once the model has been cast in the GLMM
framework, the model is fit by pseudo-restricted maximum likelihood
estimation based on linearization [\citet{Wolfinger1993}]. This
method is doubly iterative and employs Taylor series expansions to
approximate the model by a linear mixed model using pseudo-data. The
details of the estimation method are given in Appendix~B of the
Supplementary Material [\citet{supp}]. In practice, the weight smoothing models
(a)--(c) can be implemented in, for example, the procedure \textit
{GLIMMIX} in the software package SAS. See Appendix~C of the
Supplementary Material for example code [\citet{supp}].

A predictive model-based approach, by predicting the outcome of the
nonsampled individuals based on model~(\ref{eqgeneralmodel}), by
\citet{Royall1970} is used to specify an estimator of $\bar{Y}$.
The estimator is
%
\begin{equation}
\label{eqestimator} \bar{y}_{\mathit{ws}} = E(\bar{Y}|\mathbf{y}) =
\frac{1}{N} \sum_{h=1}^{H} \bigl\{
n_{h}\bar{y}_{h} + (N_{h}-n_{h})\hat{
\mu}_{h} \bigr\},
\end{equation}
where $\hat{\mu}_{h} = E(\bar{Y}_{h}|\mathbf{y}) = g^{-1}(\delta
_{h}^{*})$. The unweighted and post-stratified prevalences are special
cases of (\ref{eqestimator}), and are obtained if $\mathbf{D}
\rightarrow0$ and $\mathbf{D} \rightarrow\infty$ in (\ref
{eqgeneralmodel}), respectively.

Note that in (\ref{eqgeneralmodel}) and (\ref{eqestimator}) there is no
indication that the post-stratification weights are actually smoothed.
However, the post-stratification weights are indirectly smoothed by the
shrinkage associated with model~(\ref{eqgeneralmodel}). For Gaussian
data, explicit analytical formulas for the smoothed weights are
available [\citet{Lazzeroni1998}]. For binary data, on the
contrary, these formulas are not available due to the lack of the hat
matrix from model~(\ref{eqgeneralmodel}).

Essentially, estimator~(\ref{eqestimator}) estimates the finite
population mean by using a GLMM to predict the unobserved values. These
estimators are also in common use for prediction of small area
proportions in SAE [see, e.g., \citet{Farrell2000}].

With regard to a variance estimator of (\ref{eqestimator}), both an
analytical formula and a resampling method are presented. The
analytical approach is attractive because it only involves matrix
calculations, but it is only an approximation. As an alternative, a
more computer intensive bootstrap approach is proposed.

\subsubsection*{Analytical approach}
Because the weight smoothing model~(\ref{eqgeneralmodel}) can be cast
into a GLMM, the estimation of the variance can be derived within this
framework. To derive an approximate expression for the variance of $\bar
{y}_{\mathit{ws}}$, a first order Taylor series expansion of (\ref{eqestimator})
is taken with respect to the fixed and random effects. This leads to
%
\begin{equation}
\label{eqvariance1} \operatorname{Var}(\bar{y}_{\mathit{ws}}) \approx\frac{1}{N^{2}} (
\mathbf{N}-\mathbf{n} )^{T} \bolds{\Theta} ( \mathbf{N}-\mathbf{n}
),
\end{equation}
with $(\mathbf{N}-\mathbf{n}) = (N_{1}-n_{1},\ldots,N_{H}-n_{H})^{T}$ and
$
\bolds{\Theta} = {\bolds{\Delta}}_{*} \mathbf{C}(\mathbf{C}^{T} \bolds
{\Sigma}_{p}^{-1} \mathbf{C} + \mathbf{B} )^{-1} \mathbf{C}^{T}{\bolds
{\Delta}}_{*}.
$
Definitions of the matrices ${\bolds{\Delta}}_{*}$, $\mathbf{C}$,
$\bolds{\Sigma}_{p}$ and $\mathbf{B}$ are given in Appendix~B of the
Supplementary Material [\citet{supp}]. Finally, the variance is estimated by
replacing ${\bolds{\Delta}}_{*}$, $\bolds{\Sigma}_{p}$ and $\mathbf{B}$
by their estimates $\hat{{\bolds{\Delta}}}_{*}$, $\hat{\bolds{\Sigma
}}_{p}$ and $\hat{\mathbf{B}}$. Similar variance estimates as in (\ref
{eqvariance1}) are also obtained in model-based approaches in SAE
[\citet{Farrell2000}]. Note that the variance estimator~(\ref
{eqvariance1}) does not include a term that accounts for the
uncertainty of the estimated variance components of model~(\ref
{eqgeneralmodel}). Hence, confidence intervals based on (\ref
{eqvariance1}) are often too narrow to achieve the desired level of
coverage. A number of analytical methods for addressing this
shortcoming are available [see, e.g., \citet{Prasad1990}]. We,
however, opt to use a parametric bootstrap approach because it is a
conceptually simple alternative.

\subsubsection*{Bootstrap approach}
The described bootstrap approach was proposed in SAE [\citet
{Laird1987,Gonzalez2007}] and is easily adjusted to the case of weight
smoothing models. The approach consists of two procedures.

{\renewcommand{\theproc}{I}
\begin{proc}\label{procI}
\begin{longlist}[(I.3)]
\item[(I.1)] From the original survey, obtain estimates $\hat{\bolds{\beta}}$
and $\hat{\mathbf{D}}$ of $\bolds{\beta}$ and $\mathbf{D}$,
respectively. For the NPAR model, also obtain the empirical Bayes
prediction $\hat{\mathbf{b}}_{u}$ of the spline coefficients $\mathbf{b}_{u}$.

\item[(I.2)] Generate a random vector $\tilde{\mathbf{u}}$ from $\tilde{\mathbf
{u}} \sim\mathcal{N} ( \mathbf{0},\hat{\mathbf{D}} )$.

\item[(I.3)] Generate a population $\tilde{P}$ of size $N$ by generating
values from a binomial distribution with sizes $N_{h}$ and
probabilities $\tilde{p}_{h}$, for $h=1,\ldots,H$,\break according to the
bootstrap superpopulation model $\tilde{p}_{h} = \operatorname{expit}(\tilde
{\delta}_{h}^{*})$, where $\tilde{\bolds{\delta}}^{*} = (\tilde{\delta
}_{1}^{*},\ldots,\tilde{\delta}_{H}^{*})^{T} = \mathbf{X}\hat{\bolds
{\beta}} + \mathbf{Z}\tilde{\mathbf{u}}$ for the XRE and LIN model, and
$\tilde{\bolds{\delta}}^{*} =\break (\tilde{\delta_{1}}^{*},\ldots,\tilde
{\delta_{H}}^{*})^{T} = \mathbf{X}\hat{\bolds{\beta}} + \mathbf
{Z}_{1}\hat{\mathbf{b}}_{u} + \mathbf{Z}_{2}\tilde{\mathbf{u}}$ for the
NPAR model (see Appendix~B of the Supplementary Material [\citet{supp}]).
\end{longlist}
\end{proc}}

{\renewcommand{\theproc}{II}
\begin{proc}\label{procII}
\begin{longlist}[(II.2)]
\item[(II.1)] Use Procedure \ref{procI} to generate $B$ independent bootstrap
populations $\tilde{P}^{(b)}$ of size $N$, and calculate the bootstrap
population mean $\tilde{y}^{(b)}$.

\item[(II.2)] Extract a sample $\tilde{s}^{(b)}$ of size $n$ from each $\tilde
{P}^{(b)}$, taking into account the sample sizes in the post-strata.
Fit model~(\ref{eqgeneralmodel}) to the sample data $\tilde{s}^{(b)}$
and calculate the bootstrap predictor $\hat{\tilde{y}}^{(b)}$ using
(\ref{eqestimator}).

\item[(II.3)] The bootstrap variance of $\bar{y}_{\mathit{ws}}$ is
$
\frac{1}{B} \sum_{b=1}^{B} ( \hat{\tilde{y}}^{(b)} - \tilde{y}^{(b)} )^2
$.
\end{longlist}
\end{proc}}

From our simulation results, we recommend to take $B$ at least larger
than 100. The computing time necessary for this bootstrap method
largely depends on the computing time of the model to be estimated in
step (II.2). For the prevalence example shown in Appendix~C of the
Supplementary Material [\citet{supp}], we performed a bootstrap variance calculation
with $B=250$. For this example, step (II.2) took about 10 seconds in
total for all bootstrap samples on a laptop personal computer with
Intel Core i5-2540M processor.

\subsection{Model-assisted design-based inference}
\label{sec42}
The GREG estimator for binary outcomes, proposed by \citet
{Lehtonen1998}, is a popular model-assisted design-based estimator that
combines predicted values $\hat{y}_{i}$ based on a suitable model and
design-weighted residuals $r_{i}=y_{i}-\hat{y}_{i}$ of the sampled
units. The estimator is given by
%
\begin{equation}
\label{eqtgr} \bar{y}_{\mathit{GREG}} = \frac{1}{N} \Biggl( \sum
_{i=1}^{N} \hat{y}_{i} + \sum
_{i \in s} \frac{r_{i}}{\pi_{i}} \Biggr),
\end{equation}
where $\pi_{i}$ is the probability of inclusion in the sample for unit
$i$. The second term on the right-hand side is a bias correction factor
that eliminates possible bias due to model misspecification. \citet
{Lehtonen1998} used a design-weighted logistic regression model on
other covariates as the assisting model to predict~$\hat{y}_{i}$. Based
on estimator~(\ref{eqtgr}), we propose a model-assisted design-based
estimator $\bar{y}_{\mathit{ws},\mathit{GREG}}$ that uses a weight smoothing model as
assisting model and that can be used for observational surveys.

Define $r_{i}=y_{i}-\hat{\mu}_{h}$, where $\hat{\mu}_{h}$ is the
predicted value from (\ref{eqgeneralmodel}) of the post-stratum $h$ to
which unit $i$ belongs. The estimator $\bar{y}_{\mathit{ws},\mathit{GREG}}$ is
%
\begin{equation}
\bar{y}_{\mathit{ws},\mathit{GREG}} = \frac{1}{N} \Biggl( \sum
_{h=1}^{H} \sum_{i=1}^{N_{h}}
\hat{\mu}_{h} + \sum_{i \in s}
\frac{r_{i}}{\tilde{\pi}_{i}} \Biggr),
\end{equation}
where $\tilde{\pi}_{i}$ will be defined later. Assuming that the value
of $\tilde{\pi}_{i}$ is the same for all respondents in stratum $h$,
the estimator can be written as
%
\begin{equation}
\label{eqygreg} \bar{y}_{\mathit{ws},\mathit{GREG}} = \frac{1}{N} \sum
_{h=1}^{H} \biggl\{ \frac{n_{h}}{\tilde{\pi}_{h}}
\bar{y}_{h} + \biggl( N_{h}-\frac{n_{h}}{\tilde{\pi}_{h}} \biggr) \hat{
\mu}_{h} \biggr\}.
\end{equation}
Because we are dealing with self-selected samples in observational
surveys, no inclusion probabilities $\pi_{i}$ are available.
Additionally, setting $\tilde{\pi}_{h} = n_{h}/N_{h}$ would just return
the post-stratified mean. We propose to set the values of $\tilde{\pi
}_{h}$ using the idea of weight trimming, namely,
\begin{eqnarray*}
\tilde{\pi}_{h} &=& \frac{\hat{n}_{h}}{N_{h}}\qquad\mbox{where }
\hat{n}_{h} = \cases{ \displaystyle\frac{N_{h}/N}{w_{0}/n}, &\quad if
$w_{h} > w_{0}$,
\cr
\gamma n_{h}, &\quad if
$w_{h} \leq w_{0}$} \quad\mbox{and}
\\
\gamma&=&
\frac{n-\sum_{h: w_h>w_0}\hat{n}_{h}}{\sum_{h: w_h \leq w_0} n_{h}}.
\end{eqnarray*}
Note that for the calculation of $\gamma$, the values of $\hat{n}_{h}$
for those strata where $w_h > w_0$ need to be calculated first. In
summary, our proposed estimator $\bar{y}_{\mathit{ws},\mathit{GREG}}$ combines predicted
values based on a weight smoothing model and a weighted sum of
residuals of the sampled units where the weights $\tilde{\pi}_{h}^{-1}$
resemble trimmed weights.

For the variance estimator of $\bar{y}_{\mathit{ws},\mathit{GREG}}$, we use a jackknife
procedure described in \citet{Zheng2005} and \citet
{Chen2010}. A jackknife procedure is used because it resembles the
concept of design-based inference, in which inference is based on the
sampling distribution. The jackknife approach is given in Procedure \ref{procIII}.

{\renewcommand{\theproc}{III}
\begin{proc}\label{procIII}
\begin{longlist}[(III.3)]
\item[(III.1)] Sort the original sample $s$ according to the post-stratifying
variable $X$. Stratify the sample into $n/G$ strata, each of size $G$
with similar values of the post-stratifying variable $X$. Construct $G$
subgroups by selecting one element at a time from each stratum without
replacement.

\item[(III.2)] For each $g=1,\ldots,G$, construct the reduced sample $s^{(g)}$
from $s$ without the elements in the $g$th subgroup. Fit model~(\ref
{eqgeneralmodel}) to $s^{(g)}$ and calculate
$ \bar{y}_{\mathit{ws},\mathit{GREG}}^{(g)} $ using~(\ref{eqygreg}). Note that in this
step the weights $\tilde{\pi}_{h}^{-1}$ calculated on the original
sample $s$ are still used.

\item[(III.3)] Define $\bar{y}_{\mathit{ws},\mathit{GREG}}^{*} = G^{-1} \sum_{g=1}^{G} \bar
{y}_{\mathit{ws},\mathit{GREG}}^{(g)}$. The jackknife variance estimator of $\bar
{y}_{\mathit{ws},\mathit{GREG}}$ is
\[
\frac{G-1}{G} \sum_{g=1}^{G} \bigl(
\bar{y}_{\mathit{ws},\mathit{GREG}}^{(g)} - \bar{y}_{\mathit{ws},\mathit{GREG}}^{*}
\bigr)^2.
\]
\end{longlist}
\end{proc}}

\subsection{Extension toward trend estimation}
\label{sec43}
So far, focus was on estimating the overall prevalence. Here, a method
is proposed to estimate the overall time trend of a survey outcome
measured at different time points $t=1,\ldots,T$. Interest is in
estimating $\bar{Y}_{t} = \frac{1}{N_{t}}\sum_{i=1}^{N_{t}}Y_{it}$, for
$t=1,\ldots,T$, from a sample which is available at each time point
$t$. The sample at time~$t$ can be used to estimate $\bar{Y}_{t}$ via
the unweighted, post-stratified, trimmed or weight smoothed mean.
However, we propose to extend model~(\ref{eqgeneralmodel}) to a weight
smoothing model that exploits the time trend. The general form of the
weight smoothing model with smooth time trend is
%
\begin{eqnarray}
\label{eqgeneralmodeltime} y_{i(h),t}|\mu_{h,t} &\sim& \operatorname{Bern}(
\mu_{h,t})\quad\mbox{and}\quad \bolds{\delta^{*}} \sim
\mathcal{N}_{H}(\bolds{\delta},\mathbf{D}),
\end{eqnarray}
where $g(\mu_{h,t}) = \delta_{t} + \delta_{h}^{*}$, $\bolds{\delta
^{*}}=(\delta_{1}^{*}, \ldots, \delta_{H}^{*})^{T}$ and $\bolds{\delta}
= (\delta_{1}, \ldots, \delta_{H})^{T}$ as before. For $\bolds{\delta}$
and $\mathbf{D}$ we can again assume a XRE, LIN or NPAR model as in
Section~\ref{sec41}. The parameter $\delta_{t}$ corresponds to the time
trend and is modeled by a nonparametric function, $f_{t}(\cdot)$, which is
specified by the approximating thin plate spline family. Model~(\ref
{eqgeneralmodeltime}) can again be cast in the GLMM framework. After
model fitting, the fitted post-strata means, $\hat{\bolds{\mu
}}_{t}=(\hat{\mu}_{1,t},\ldots,\hat{\mu}_{H,t})^T$, are obtained at
each time point $t$, and an estimator of $\bar{Y}_{t}$ is
%
\begin{equation}
\label{eqytime} \bar{y}_{\mathit{ws},t} = \frac{1}{N_{t}} \sum
_{h=1}^{H} \bigl\{ n_{h,t}
\bar{y}_{h,t} + (N_{h,t}-n_{h,t})\hat{
\mu}_{h,t} \bigr\}.
\end{equation}
A GREG adjusted estimator, $\bar{y}_{\mathit{ws},t,\mathit{GREG}}$, is constructed in the
same manner as in Section~\ref{sec42},
%
\begin{equation}
\label{eqygregtime} \bar{y}_{\mathit{ws},t,\mathit{GREG}} = \frac{1}{N_{t}} \sum
_{h=1}^{H} \biggl\{ \frac{n_{h,t}}{\tilde{\pi}_{h,t}}
\bar{y}_{h,t} + \biggl(N_{h,t}-\frac{n_{h,t}}{\tilde{\pi}_{h,t}}\biggr
)\hat{
\mu}_{h,t} \biggr\}.
\end{equation}

Variance estimation of $\bar{y}_{\mathit{ws},t}$ can be obtained by an
analytical formula or a bootstrap approach. The approximate analytical
variance is given by
%
\begin{equation}
\label{eqvariancetrendestimation}
\operatorname{Var}({\bar{y}}_{\mathit{ws},t}) \approx
\frac{1}{N_{t}^{2}} (
\mathbf{N}_{t}-\mathbf{n}_{t})^{T} \bolds{
\Theta}_{t} (\mathbf{N}_{t}-\mathbf{n}_{t}),
\end{equation}
where $(\mathbf{N}_{t}-\mathbf{n}_{t})=(N_{1,t}-n_{1,t},\ldots
,N_{H,t}-n_{H,t})^T$ and $\bolds{\Theta}_{t}$, for $t=1,\ldots,T$, are
the subsequent $H \times H$ block matrices along the main diagonal of
the $HT \times HT$ covariance matrix $\bolds{\Theta}$ of the fitted
post-stratum means, calculated, as before, by a first order Taylor
series expansion. For the bootstrap approach, Procedure \ref{procI} and Procedure
\ref{procII} proceed analogously with the adjustment that bootstrap data is now
sampled from model~(\ref{eqgeneralmodeltime}).

%
\begin{table}[t]
\tabcolsep=0pt
\caption{Population and sample sizes in the 18 strata for the large
population ($N^{(1)}={}$6,000,000) and small population
($N^{(2)}={}$150,000) used in the simulation study.}\label{Table1}
\begin{tabular*}{\tablewidth}{@{\extracolsep{\fill}}@{}ld{6.0}d{6.0}d{6.0}d{6.0}d{6.0}d{6.0}d{6.0}d{6.0}d{6.0}@{\hspace*{-2pt}}}
\hline
\textbf{Stratum $\bolds{h}$} & \multicolumn{1}{c}{\textbf{1}} & \multicolumn{1}{c}{\textbf{2}} & \multicolumn{1}{c}{\textbf{3}}
& \multicolumn{1}{c}{\textbf{4}} & \multicolumn{1}{c}{\textbf{5}} & \multicolumn{1}{c}{\textbf{6}} & \multicolumn{1}{c}{\textbf{7}}
& \multicolumn{1}{c}{\textbf{8}} & \multicolumn{1}{c@{}}{\textbf{9}} \\
\hline
$N_{h}^{(1)}$ & 300{,}000 & 300{,}000 & 320{,}000 & 320{,}000 & 340{,}000 & 340{,}000 & 360{,}000 & 360{,}000 & 360{,}000\\[2pt]
$n_{1,h}^{(1)}$ & 50& 150& 300& 750& 1250& 1500& 2000& 2750& 3750\\[2pt]
$n_{2,h}^{(1)}$ & 10& 30& 60& 150& 250& 300& 400& 550& 750\\[2pt]
$N_{h}^{(2)}$ & 7500 & 7500 & 8000 & 8000 & 8500 & 8500 & 9000 & 9000 & 9000\\[2pt]
$n_{1,h}^{(2)}$ & 5& 15& 30& 75& 125& 150& 200& 275& 375\\[2pt]
$n_{2,h}^{(2)}$ & 1& 3& 6& 15& 25& 30& 40& 55& 75\\[12pt]
\hline
\textbf{Stratum $\bolds{h}$} & \multicolumn{1}{c}{\textbf{10}} & \multicolumn{1}{c}{\textbf{11}} & \multicolumn{1}{c}{\textbf{12}}
& \multicolumn{1}{c}{\textbf{13}} & \multicolumn{1}{c}{\textbf{14}} & \multicolumn{1}{c}{\textbf{15}}
& \multicolumn{1}{c}{\textbf{16}} & \multicolumn{1}{c}{\textbf{17}} & \multicolumn{1}{c@{}}{\textbf{18}}\\
\hline
$N_{h}^{(1)}$ & 360{,}000 & 360{,}000 & 360{,}000 & 340{,}000 & 340{,}000 & 320{,}000 & 320{,}000 & 300{,}000 & 300{,}000\\[2pt]
$n_{1,h}^{(1)}$ & 3750& 2750& 2000& 1500& 1000& 800& 400& 200& 100\\[2pt]
$n_{2,h}^{(1)}$ & 750& 550& 400& 300& 200& 160& 80& 40& 20\\[2pt]
$N_{h}^{(2)}$ & 9000 & 9000 & 9000 & 8500 & 8500 & 8000 & 8000 & 7500 & 7500\\[2pt]
$n_{1,h}^{(2)}$ & 375& 275& 200& 150& 100& 80& 40& 20& 10\\[2pt]
$n_{2,h}^{(2)}$ & 75& 55& 40& 30& 20& 16& 8& 4& 2\\
\hline
\end{tabular*}
\end{table}

For the jackknife variance calculation of $\bar{y}_{\mathit{ws},t,\mathit{GREG}}$, some
technical adjustments need to be made to Procedure \ref{procIII}. The jackknife
procedure is provided in Procedure~\ref{procIV}.

{\renewcommand{\theproc}{IV}
\begin{proc}\label{procIV}
\begin{longlist}[(IV.3)]
\item[(IV.1)] Sort the original sample $s$ first according to time and second
to the post-stratifying variable $X$. At each time point, stratify the
sample into $n_{t}/G$ strata each of size $G$ with similar values of
the post-stratifying variable $X$. Construct $G$ subgroups by selecting
one element at a time from each stratum without replacement.

\item[(IV.2)] For each $g=1,\ldots,G$, construct the reduced sample $s^{(g)}$
from $s$ where at each time point the elements in the $g$th subgroup
are removed. Fit model~(\ref{eqgeneralmodeltime}) to $s^{(g)}$ and calculate
$ \bar{y}_{\mathit{ws},t,\mathit{GREG}}^{(g)} $ using (\ref{eqygregtime}).\vspace*{1pt}

\item[(IV.3)] Similar as in (III.3).
\end{longlist}
\end{proc}}

\section{Simulation study}
\label{sec5}
\subsection{Simulation settings for overall prevalence estimation}
\label{sec51}
In total, $8 \times2 \times2= 32$ simulation conditions are evaluated
for the estimation of the\vadjust{\goodbreak} overall prevalence. This is done by crossing
eight population types with two population sizes and two sample sizes
for each population size. For the first population a total size
$N^{(1)}={}$6,000,000 is considered and for the second size
$N^{(2)}={}$150,000. It is assumed that both populations consist of 18
strata with strata population sizes given in Table~\ref{Table1}.

The population values are generated as in model~(\ref{eqgeneralmodel})
with the following choices:

\begin{longlist}[(2)]
\item[(1)] NULL: $\delta_{h}=0$ ($\forall h=1,\ldots,18$) and $\mathbf {D}=\mathbf{0}$.

\item[(2)] XRE: $\delta_{h}=0$ ($\forall h=1,\ldots,18$) and $\mathbf {D}=\sigma^{2} \mathbf{I}$ with $\sigma^{2}=0.02$.

\item[(3)] LIN$_{0}$: $\delta_{h}=-2+0.2h$ ($\forall h=1,\ldots,18$) and $\mathbf{D}=\mathbf{0}$.

\item[(4)] LIN$_{1}$: $\delta_{h}=-2+0.2h$ ($\forall h=1,\ldots,18$) and $\mathbf{D}=\sigma^{2} \mathbf{I}$ with $\sigma^{2}=0.02$.

\item[(5)] QUAD$_{0}$: $\delta_{h}=1-0.25h+0.01h^{2}$ ($\forall h=1,\ldots ,18$) and $\mathbf{D}=\mathbf{0}$.

\item[(6)] QUAD$_{1}$: $\delta_{h}=1-0.25h+0.01h^{2}$ ($\forall h=1,\ldots ,18$) and $\mathbf{D}=\sigma^{2} \mathbf{I}$ with $\sigma^{2}=0.02$.

\item[(7)] EXP$_{0}$: $\delta_{h}=-1+2\exp ( -\frac{h}{9} )$ ($\forall h=1,\ldots,18$) and $\mathbf{D}=\mathbf{0}$.

\item[(8)] EXP$_{1}$: $\delta_{h}=-1+2\exp ( -\frac{h}{9} )$ ($\forall
h=1,\ldots,18$) and $\mathbf{D}=\sigma^{2} \mathbf{I}$ with $\sigma^{2}=0.02$.
\end{longlist}

The parameters are chosen such that the overall mean
prevalence in the population is about 0.5. For each of the above
population models (1)--(8), 25 populations are randomly generated and
this for both population sizes $N^{(1)}$ and $N^{(2)}$. In each of
these populations, ten samples of fixed sample size are generated. This
procedure yields a total of 250 replications for each combination of
population size, sample size and population model. For the large
population size ($N^{(1)}$) we consider samples of size 25,000 and size
5000. For the small population size ($N^{(2)}$) we consider samples of
sizes 2500 and 500, respectively. The sample sizes per stratum are
given in Table \ref{Table1}. In all of the settings the normalized
post-stratification weights of the strata are similar, ranging from 0.4
to 25.

The following nine estimators for estimating the overall population
mean from the generated samples are compared: (\textit{psm}) the
post-stratified mean, $\bar{y}_{\mathit{ps}}$; (\textit{unw}) the unweighted
mean, $\bar{y}_{\mathit{unw}}$; (\textit{trim}) the trimmed mean, $\bar
{y}_{\mathit{trim}}$, with a cutoff value of $w_{0}=3$; (\textit{xre}) $\bar
{y}_{\mathit{ws}}$, using the XRE assumption; (\textit{xre-greg}) $\bar
{y}_{\mathit{ws},\mathit{greg}}$, using the XRE assumption; (\textit{lin}) $\bar
{y}_{\mathit{ws}}$, using the LIN assumption; (\textit{lin-greg}) $\bar
{y}_{\mathit{ws},\mathit{greg}}$, using the LIN assumption; (\textit{npar}) $\bar
{y}_{\mathit{ws}}$, using the NPAR assumption; (\textit{npar-greg}) $\bar
{y}_{\mathit{ws},\mathit{greg}}$, using the NPAR assumption. For all GREG estimators we
use $w_{0}=3$. For the \textit{npar} estimator we place knots at each
value of $h$, for a total of 18 knots. A sensitivity analysis with ten
equally spaced knots revealed no meaningful differences.

For each of these estimators we calculate the average bias,
variability, mean squared error (MSE), the coverage and average length
of the 95$\%$ confidence interval (CI). To calculate the MSE, we first
calculate the mean squared error within each of the 25 populations.
Denote by $\theta^{(p)}$ the true population proportion of population
$p$. For each population,\vspace*{1pt} we obtain ten estimates $\hat{\bolds{\theta
}}^{(p)} = (\hat{\theta}^{(p)}_{1},\ldots,\hat{\theta}^{(p)}_{10} )$ of
$\theta^{(p)}$. The mean squared error in population $p$ is estimated
by MSE$^{(p)} = \operatorname{Var}(\hat{\bolds{\theta}}^{(p)}) +  (\operatorname{Bias}(\theta^{(p)}, \hat{\bolds{\theta}}^{(p)}) )^{2}$. The overall
mean squared error is then calculated by averaging over the 25
MSE$^{(p)}$ values.
For the bootstrap and jackknife variance procedure we used $B=250$ and
$G=250$, respectively. The CIs are first calculated on the linear
scale. Next, these CIs are back-transformed to yield CIs between 0 and
1. A normal reference distribution is used for the CIs. For \textit
{psm}, \textit{unw} and \textit{trim} we use the variance formulas as
given in Table~1 in \citet{Little1991}.

\begin{table}
\tabcolsep=0pt
\caption{Mean squared error ($\times1$0$^4$) of nine estimators and
eight population models for the large population $N^{(1)}={}$6,000,000
with a sample size of 5000. $^{\dagger}$Based on 249 simulated data sets}\label{Table2}
\begin{tabular*}{\tablewidth}{@{\extracolsep{\fill}}@{}lccccd{2.2}d{2.2}cc@{}}
\hline
\textbf{Estimator} & \textbf{NULL} & \textbf{XRE} & \textbf{LIN}$_{\bolds{0}}^{\bolds{\dagger}}$
& \textbf{LIN}$_{\bolds{1}}$ & \multicolumn{1}{c}{\textbf{QUAD}$_{\bolds{0}}$} & \multicolumn{1}{c}{\textbf{QUAD}$_{\bolds{1}}$}
& \textbf{EXP}$_{\bolds{0}}$ & \textbf{EXP}$_{\bolds{1}}$\\
\hline
\textit{psm} & 1.70 & 2.85 & 1.16 & 1.73 & 1.88 & 2.92 & 1.82 & 2.90\\
\textit{unw} & 0.46 & 1.83 & 0.43 & 1.54 & 16.09 & 17.52 & 4.16 & 5.42\\
\textit{trim} & 0.68 & 1.92 & 0.86 & 1.68 & 5.61 & 6.50 & 2.41 & 3.26\\
\textit{xre} & 0.47 & 0.61 & 1.01 & 1.04 & 4.64 & 4.11 & 1.83 & 1.75\\
\textit{xre-greg} & 0.67 & 0.84 & 1.06 & 1.09 & 3.14 & 2.90 & 1.73 & 1.68\\
\textit{lin} & 0.48 & 0.61 & 0.30 & 0.35 & 8.35 & 5.59 & 2.97 & 2.04\\
\textit{lin-greg} & 0.67 & 0.83 & 0.50 & 0.53 & 3.66 & 2.96 & 1.49 & 1.32\\
\textit{npar} & 0.65 & 0.90 & 0.48 & 0.57 & 1.63 & 1.71 & 1.74 & 1.49\\
\textit{npar-greg} & 0.85 & 1.08 & 0.66 & 0.71 & 1.52 & 1.57 & 1.49 & 1.40\\
\hline
\end{tabular*}
\end{table}
%
\begin{figure}

\includegraphics{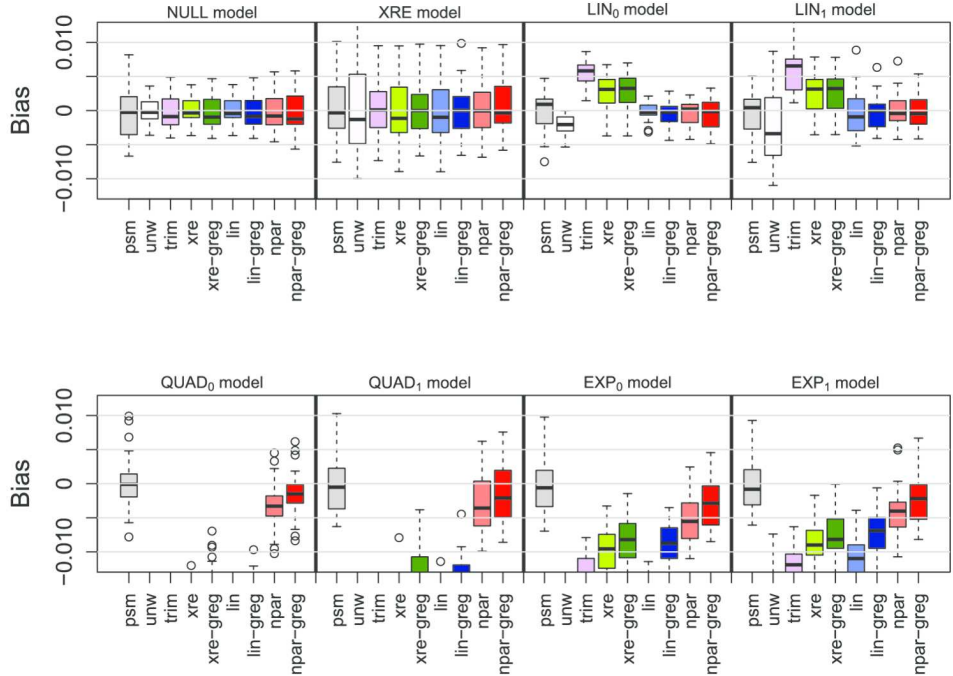}

\caption{Bias results of nine estimators and eight population models
for the large population $N^{(1)}={}$6,000,000 with a sample size of
5000. Boxplots of the average bias associated with the 25 simulated
populations are shown. The bias of some estimators exceeds the range of
the y-axis used and are therefore not depicted on the figure.}
\label{Figure2}
\end{figure}
%

\subsection{Results for overall prevalence estimation}
\label{sec52}
Only part of the simulation results are presented here. For more
results we refer to Appendix~D of the Supplementary Material [\citet{supp}].
Table~\ref{Table2} summarizes the MSE and Figure~\ref{Figure2} shows
the bias for $N^{(1)}$ with a sample size of 5000.
When the underlying population has a constant mean (NULL and XRE), all
estimators yield unbiased estimates. For the QUAD and EXP populations,
the GREG adjusted estimators show less bias than their nonadjusted
counterparts. The \textit{psm} estimator remains unbiased in all
simulation settings. In general, \textit{psm} does not perform well in
terms of MSE, due to the overwhelming increase in variability over the
other estimators (see Appendix~D of the Supplementary Material [\citet{supp}]). For a
population with more structure (LIN, QUAD and EXP), the \textit{unw}
and \textit{trim} estimators show a high MSE, due to the bias of the
estimates (see Figure~\ref{Figure2}). For the LIN populations, as
expected, the estimators \textit{lin} and \textit{lin-greg} performed
the best in terms of MSE. The \textit{npar} estimator is second best
with a slight increase in MSE. The \textit{npar} estimators perform the
best for the QUAD populations. For the EXP populations, the \textit
{lin-greg} and both \textit{npar} estimators perform well. The \textit
{npar} estimator has a higher MSE than \textit{xre} and \textit{lin}
for the NULL and XRE populations. This is due to the higher variance
associated with the \textit{npar} estimator. For the QUAD and EXP
underlying populations, it is observed that the estimators \textit
{xre-greg}, \textit{lin-greg} and \textit{npar-greg} have lower MSE
when compared with the nonadjusted estimators. In the QUAD and EXP
scenarios, the \textit{npar-greg} yields smaller MSE values than the
\textit{npar}, which could suggest that the NPAR weight smoothing model
does not fit the data generated by the QUAD and EXP models well.
Investigating the model fits reveals no problems with the fits of the
NPAR weight smoothing model (see Appendix~E of the Supplementary
Material [\citet{supp}]). The bias reduction of the \textit{npar-greg} achieved by
the bias reduction term is greater than the increase in variability of
this estimator, which leads to smaller MSE values for the \textit
{npar-greg} in the QUAD and EXP scenarios. Overall, the \textit{npar}
and \textit{npar-greg} estimators perform the most consistent in the
simulation study. Results for other population and sample sizes do not
differ qualitatively (see Appendix~D of the Supplementary Material [\citet{supp}]).

\begin{sidewaystable}
\tabcolsep=0pt
\tablewidth=\textwidth
\caption{Nominal\vspace*{1pt} coverage of the 95$\%$ CI (average length $\times$100
of the CI) of nine estimators for the large population
[$N^{(1)}={}$6,000,000] with a sample size~of~5000}\label{Table3}
\begin{tabular*}{\tablewidth}{@{\extracolsep{\fill}}@{}lcccccccc@{}}
\hline
\textbf{Estimator} & \textbf{NULL} & \textbf{XRE} & \textbf{LIN}$_{\bolds{0}}^{\bolds{\dagger}}$ & \textbf{LIN}$_{\bolds{1}}$
& \textbf{QUAD}$_{\bolds{0}}$ & \textbf{QUAD}$_{\bolds{1}}$ & \textbf{EXP}$_{\bolds{0}}$ & \textbf{EXP}$_{\bolds{1}}$\\
\hline
\textit{psm} & 95.2 (5.3) & 95.2 (5.3) & 93.6 (4.1) & 92.8 (4.1) & 93.2 (5.1) & 94.8 (5.1) & 93.6 (5.0) & 92.4 (5.0)\\
\textit{unw} & 96.8 (2.8) & 87.2 (2.8) & 95.6 (2.7) & 88.8 (2.7) & \phantom{0}0.0 (2.7) & \phantom{0}0.0 (2.7) & 22.0 (2.7) & 26.4 (2.7)\\
\textit{trim} & 95.6 (3.3) & 92.8 (3.3) & 88.4 (2.9) & 84.4 (2.9) & 24.0 (3.2) & 28.0 (3.2) & 64.4 (3.2) & 65.2 (3.2)\\
\textit{xre} (analytical) & 96.8 (2.8) & 94.0 (3.2) & 96.8 (4.2) & 96.4 (4.2) & 51.2 (3.9) & 58.0 (4.0) & 87.2 (4.1) & 86.8 (4.1)\\
\textit{xre} (bootstrap) & 97.2 (2.9) & 94.0 (3.3) & 98.8 (4.5) & 98.8 (4.5) & 50.4 (3.9) & 58.0 (4.0) & 86.4 (4.1) & 88.8 (4.1)\\
\textit{xre-greg} (jackknife) & 98.0 (3.4) & 96.0 (3.7) & 96.4 (3.9) & 95.6 (3.9) & 58.8 (4.5) & 64.0 (4.5) & 85.6 (4.0) & 84.4 (4.2)\\
\textit{lin} (analytical) & 96.8 (2.8) & 94.8 (3.2) & 96.8 (2.4) & 95.2 (2.6) & 19.6 (3.2) & 36.0 (3.5) & 40.4 (2.8) & 69.2 (3.2)\\
\textit{lin} (bootstrap) & 97.2 (2.9) & 94.4 (3.3) & 96.8 (2.4) & 95.6 (2.6) & 19.2 (3.2) & 37.2 (3.5) & 41.2 (2.8) & 69.6 (3.2)\\
\textit{lin-greg} (jackknife) & 98.0 (3.4) & 96.8 (3.5) & 98.8 (2.9) & 96.4 (2.9) & 25.6 (3.8) & 42.4 (3.9) & 60.0 (3.3) & 73.2 (3.5)\\
\textit{npar} (analytical) & 94.8 (3.0) & 93.2 (3.5) & 95.6 (2.6) & 94.0 (2.8) & 91.2 (4.4) & 89.2 (4.4) & 77.6 (3.7) & 86.4 (3.8)\\
\textit{npar} (bootstrap) & 97.6 (3.6) & 96.0 (3.6) & 98.4 (2.9) & 95.2 (3.0) & 94.4 (4.9) & 92.0 (4.8) & 84.8 (4.2) & 87.6 (4.1)\\
\textit{npar-greg} (jackknife) & 98.0 (4.0) & 96.8 (4.1) & 98.8 (3.4) & 96.8 (3.4) & 93.2 (4.6) & 92.4 (4.7) & 87.6 (4.4) & 90.0 (4.4)\\
\hline
\end{tabular*}
\tabnotetext[]{TT1}{$^{\bolds{\dagger}}$ Based on 249 simulated data sets.}
\end{sidewaystable}
Table~\ref{Table3} shows the nominal coverage and average length of the
95$\%$ CI for $N^{(1)}$ with a sample size of 5000. The \textit{psm}
estimator attains a good coverage over all simulation conditions. Due
to the bias, the nominal coverage of the \textit{unw} and \textit{trim}
estimators deteriorates when the underlying model has more structure.
For the underlying models with nonlinear mean (QUAD and EXP), the
\textit{xre} and \textit{lin} estimators do not obtain good coverage
results. Overall, the \textit{npar-greg} estimator yields consistent
nominal coverage results over all the simulation conditions. Only for
the QUAD and EXP populations, the \textit{npar-greg} underestimates the
actual coverage slightly. The average length of the CIs is also smaller
for the \textit{npar-greg} method when compared to \textit{psm}.

\subsection{Simulation settings for mean trend estimation}
\label{sec53}
For trend estimation, $2 \times2 \times4 = 16$ simulation conditions
are considered. This is done by crossing four population types with the
same population and sample sizes as in Section~\ref{sec51}. We assume
that the survey is available at 30 time points which are equally spaced
in time. Further, it is assumed that the post-strata population and
sample sizes remain constant over time. Similarly, as in Section~\ref
{sec51}, 25 populations are randomly generated and ten samples are
generated in each population. The population values are generated
according to model~(\ref{eqgeneralmodeltime}) with the following
choices ($\forall h=1,\ldots,18$ and $\forall t=1,\ldots,30$):

\begin{longlist}[F3:]
\item[F1:] $\delta_{h} = -1 + 2 \exp ( -\frac{h}{9} ) $, $\delta_{t} = -2 +
3\exp ( -\frac{(t-15)^2}{50} )$ and $\mathbf{D}=\mathbf{0}$.

\item[F2:] $\delta_{h} = -1 + 2 \exp ( -\frac{h}{9} ) $, $\delta_{t} = -2 +
3\exp ( -\frac{(t-15)^2}{50} )$ and $\mathbf{D}=\sigma^2 \mathbf{I}$
with $\sigma^2=0.02$.

\item[F3:] $\delta_{h} = -1 + 2 \exp ( -\frac{h}{9} ) $, $\delta_{t} = -2 +
3\exp ( -\frac{(t-15)^2}{50} ) - \exp ( -(t-15)^2 )$ and $\mathbf
{D}=\mathbf{0}$.

\item[F4:] $\delta_{h} = -1 + 2 \exp ( -\frac{h}{9} ) $, $\delta_{t} = -2 +
3\exp ( -\frac{(t-15)^2}{50} ) - \exp ( -(t-15)^2 )$ and $\mathbf
{D}=\sigma^2 \mathbf{I}$ with $\sigma^2=0.02$.
\end{longlist}

\begin{figure}[b]

\includegraphics{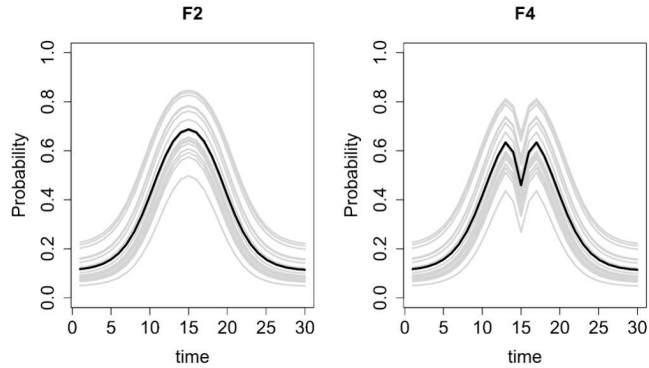}

\caption{Mean time trends of the separate post-strata (grey lines) and
total population (black line) for a random population generated
according to population models \textup{F2} and \textup{F4}.}
\label{Figure3}
\end{figure}

Figure \ref{Figure3} shows the time trends of a population
that is randomly generated from F2 and F4. We consider population types
F3 and F4 to investigate the robustness of the different estimation
methods to a sharp valley in the overall trend. The nonparametric time
function, $f_{t}(\cdot)$, is not able to perfectly describe this valley. We
shall investigate whether the GREG adjusted estimators offer a useful
solution in such case.

The following five estimators are used to obtain the time trend in the
simulation study: (\textit{psm}) the post-stratified mean is calculated
at each time point; (\textit{unw}) the unweighted mean is calculated at
each time point; (\textit{trim}) the trimmed mean \mbox{($w_0=3$)} is
calculated at each time point; (\textit{npar}) estimator~(\ref
{eqytime}) is used where the NPAR model is assumed for $\bolds{\delta}$
and $\mathbf{D}$; (\textit{npar-greg}) estimator~(\ref{eqygregtime})
with $w_0=3$ is used with a NPAR model assumption for $\bolds{\delta}$
and~$\mathbf{D}$.

For each of the above five estimators, we calculate the average bias,
variability and mean squared error from the obtained estimates at each
of the 30 time points. For each time point, this is done in a similar
manner as done for the prevalence estimation described in Section~\ref
{sec51}. The nominal coverage and average length of the 95$\%$
point-wise confidence interval are calculated. For the nonparametric
time function, we set knots at all time points to ensure enough flexibility.

\subsection{Results for mean trend estimation}
\label{sec54}
%
\begin{figure}[b]

\includegraphics{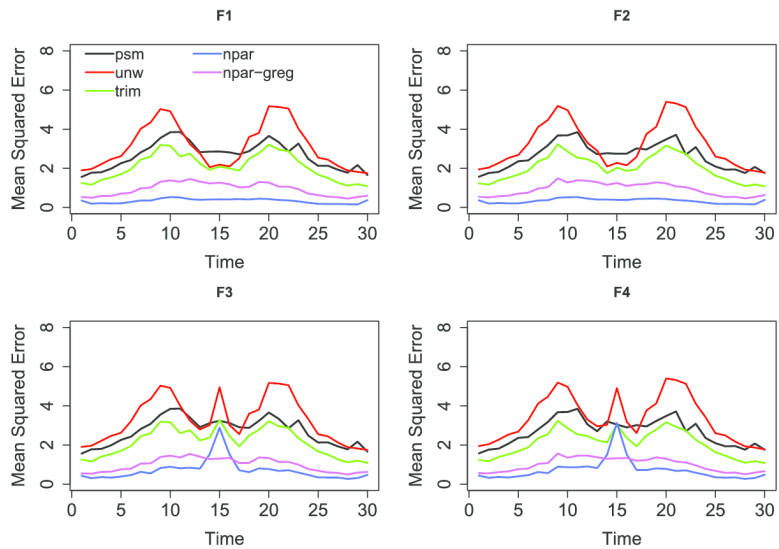}

\caption{Mean squared errors $\times$10$^4$ at each time point for
population models \textup{F1}, \textup{F2}, \textup{F3} and \textup{F4} with population size
$N^{(2)}=150{,}000$ and sample size 2500.}
\label{Figure4}
\end{figure}

A part of the results is summarized in Figure~\ref{Figure4}. The mean
squared errors $\times$10$^4$ are presented for a population size of
150,000 and a sample size of 2500. More figures on the simulation
study are given in Appendix~D of the Supplementary Material [\citet{supp}]. The
\textit{unw} and \textit{trim} estimators are biased for all population
models. The \textit{npar} estimators are unbiased for F1 and F2. For F3
and F4, however, they are unbiased at the start and end of the time
trend, but show some severe bias in the region of the sharp valley. No
bias is present for the \textit{psm} and \textit{npar-greg} estimators.
The variance of \textit{psm} is larger when compared to the other
estimators. The \textit{npar-greg} estimator exhibits more variability
than \textit{unw} and \textit{npar}, and has a comparable variability
as the \textit{trim} estimator.

In terms of MSE, this leads to a better performance of \textit{npar}
and \textit{npar-greg} when compared to the other methods. For all time
points, the \textit{npar} estimator has lower MSE values for F1 and F2
when compared with \textit{npar-greg}. For the functions with the sharp
valley, F3 and F4, the mean squared error of \textit{npar} is smaller
than the mean squared error of \textit{npar-greg}, except at those time
points where the sharp valley occurs. The MSE of the \textit{npar}
estimator shows a steep increase in that region. Thus, for cases such
as scenarios F3 and F4, the \textit{npar} estimator is not preferred,
but the \textit{npar-greg} estimator is recommended. Under varying
simulation conditions --- increasing and decreasing population and sample
sizes --- the same general results are observed.

\begin{figure}

\includegraphics{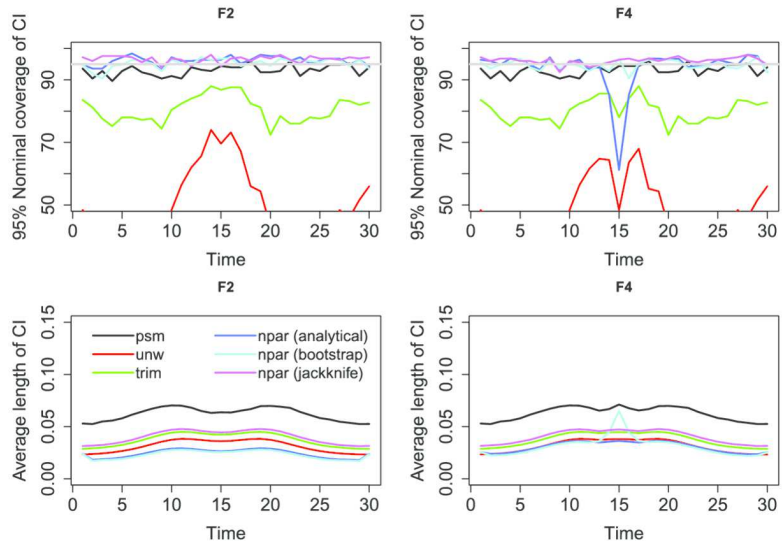}

\caption{Nominal coverage and average length of the 95$\%$ point-wise
confidence intervals at each time point for population models \textup{F2} and \textup{F4}
with population size $N^{(2)}=150{,}000$ and sample size 2500. The
nominal coverage of some estimators is smaller than the range of the
$y$-axis used and are therefore not depicted on the figure.}\label{Figure5}
\end{figure}
Results of the nominal coverage and average length of the 95$\%$
point-wise confidence intervals are presented in Figure~\ref{Figure5}
(see Appendix~D for additional results). The \textit{unw} and \textit
{trim} estimators have very poor coverage resulting from the bias
associated with the estimates. The \textit{psm} slightly underestimates
the true coverage at almost all time points.
For population model F2, both the analytical and bootstrap based CIs of
the \textit{npar} estimator have good coverage values. For F4, the
coverage of the \textit{npar} CIs based on the analytical variance
shows a drop in the sharp valley area. The bootstrap variance-based CIs
attain a good nominal coverage. However, for F4 this comes at the cost
of wider CIs in the sharp valley area. The jackknife based CIs of the
\textit{npar-greg} estimator achieve good coverage for both F2 and F4
for all time points. The average lengths of the CIs of the \textit
{npar-greg} are comparable with those of \textit{trim} and are
substantially smaller when compared to \textit{psm}.

\section{Motivating example revisited}
\label{sec6}
%
\begin{table}[t]
\tabcolsep=0pt
\caption{Estimates and 95$\%$ confidence intervals (CI) of the overall
prevalence of ILI during the 2010--2011 influenza season based on nine
estimators.}\label{Table4}
\begin{tabular*}{\tablewidth}{@{\extracolsep{\fill}}@{}lcccc@{}}
\hline
\textbf{Method} & \textbf{Estimate} & \textbf{Analytical CI} & \textbf{Bootstrap CI} & \textbf{Jackknife CI}\\
\hline
\textit{psm} & 7.10$\%$ & [5.31--9.45$\%$] & -- & --\\
\textit{unw} & 5.12$\%$ & [4.52--5.80$\%$] & -- & --\\
\textit{trim} & 5.61$\%$ & [4.88--6.46$\%$] & -- & --\\
\textit{xre} & 5.96$\%$ & [4.95--7.17$\%$] & [5.00--7.10$\%$] &-- \\
\textit{xre-greg} & 6.16$\%$ & -- & -- & [4.92--7.68$\%$] \\
\textit{lin} & 6.88$\%$ & [5.69--8.30$\%$] & [5.76--8.21$\%$] & -- \\
\textit{lin-greg} & 6.82$\%$ & -- & -- & [5.66--8.20$\%$] \\
\textit{npar} & 6.88$\%$ & [5.69--8.30$\%$] & [5.61--8.41$\%$] & -- \\
\textit{npar-greg} & 6.82$\%$ & -- & -- & [5.66--8.20$\%$]\\
\hline
\end{tabular*}
\end{table}

We apply the methods developed in Section~\ref{sec4} to the Great
Influenza Survey of 2010--2011 introduced in Section~\ref{sec2}. The
same nine estimators as in Section~\ref{sec51} are used to calculate
the overall ILI prevalence. The results are presented in Table~\ref
{Table4}. To calculate the CIs for \textit{psm}, \textit{unw} and
\textit{trim}, we use the variance formulas given in Table 1 in
\citet{Little1991}. Both the analytical and bootstrap (with
$B=250$) based CIs are calculated for \textit{xre}, \textit{lin} and
\textit{npar}. The jackknife procedure (with $G=250$) is used for the
variance of the GREG-adjusted estimates.

The post-stratified mean yields the largest estimated prevalence. This
is because the younger age groups, which have the highest ILI
prevalence, receive high post-stratification weights. The \textit{xre}
estimate yields an estimate in between the post-stratified and
unweighted mean. The difference between the \textit{xre} and \textit
{xre-greg} estimates is larger than the difference between the
estimates of \textit{lin} and \textit{lin-greg}, and \textit{npar} and
\textit{npar-greg}. This indicates that the XRE model assumption is
most likely a misspecification for this data. The point estimates
\textit{lin} and \textit{npar} are similar because the nonparametric
function in the NPAR model is estimated to be essentially linear. The
point estimates based on the LIN and NPAR model assumptions are closer
to the post-stratified mean than to the unweighted and trimmed mean.
The $\sigma^2$ parameter was estimated to be 0.068 in the LIN and NPAR
models. The post-stratified mean has the widest confidence interval
associated with its point estimate. Confidence intervals of \textit
{lin} and \textit{npar} are less wide. Reductions of 30--40$\%$ are
observed when compared with psm.

\begin{figure}

\includegraphics{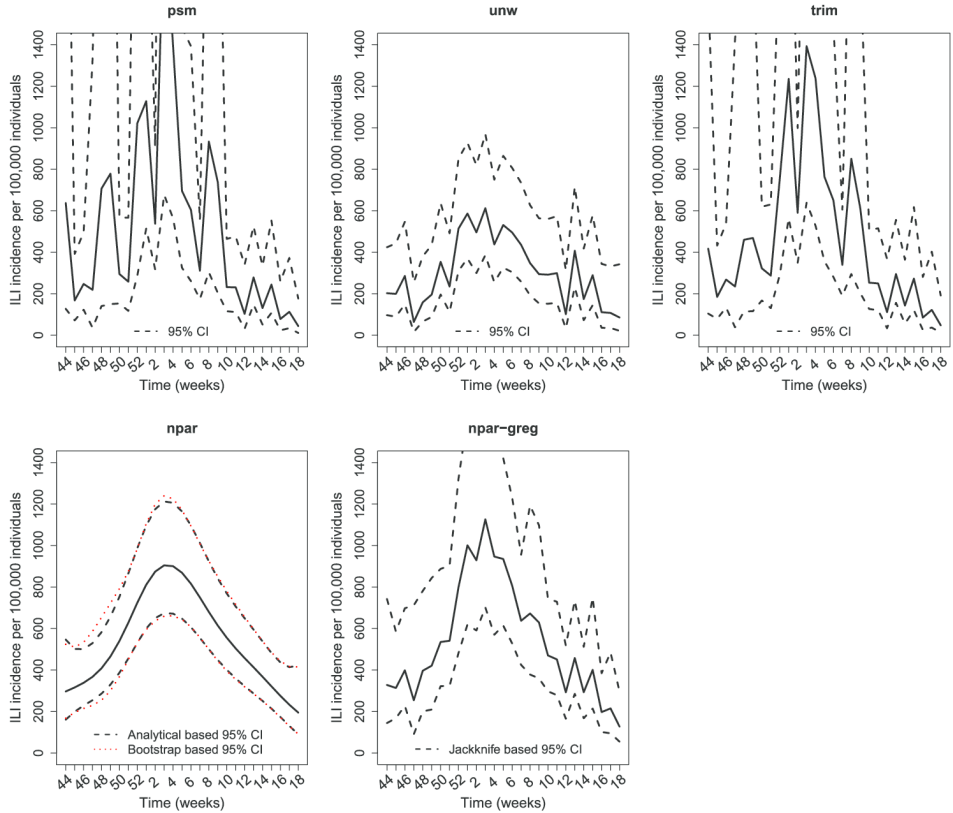}

\caption{Estimated incidence trends with corresponding 95$\%$
point-wise confidence intervals from the 2010/2011 Great Influenza
Survey (GIS) using five estimation methods.}
\label{Figure6}
\end{figure}

The same five estimators as in Section~\ref{sec53} are used to obtain
the overall time trend (the time trend represents the ILI incidence in
this data example). We take both $B=250$ and $G=250$ for the bootstrap
and jackknife procedures. Twenty equally spaced knots are used for the
nonparametric time function. A sensitivity analysis with 10 knots and
knots at all 27 time points revealed no meaningful differences.
Figure~\ref{Figure6} presents the estimated trends with the
corresponding point-wise 95$\%$ confidence intervals. It is observed
that the post-stratified mean yields a very wiggly curve. This trend
estimate is not useful for practical usage. The unweighted and trimmed
mean trend are more stable but most likely produce biased estimates as
was observed in the simulation study. The \textit{npar-greg} yields a
trend that is less smooth than \textit{npar}. The width of the
confidence intervals of both \textit{npar} and \textit{npar-greg} yield
noticeable reductions in length when compared to the post-stratified
mean trend.
%

\section{Discussion}
\label{sec7}
In this article we examined methods that can deal with ordinal
post-stratifiers that yield high post-stratification weights in
observational survey data. Standard methods such as the
post-stratified, unweighted and trimmed mean break down in this
situation due to either substantial bias or high variability of the
obtained estimates. Weight smoothing models modify the standard
methodology by imposing a random-effects structure on the post-stratum
means. By predicting the unobserved values based on the weight
smoothing model, an estimate is obtained. The post-stratification
weights are implicitly smoothed in this manner. We described two
extensions on the existing literature of weight smoothing models for
binary data. First, we proposed a GREG-adjusted weight smoothed
estimator that can be used for observational data for which no
inclusion probabilities are available. Second, an extension for trend
estimation was considered. XRE, LIN and NPAR model assumptions for the
weight smoothing models were considered.\looseness=1

The construction of the weights $\tilde{\pi}_{h}^{-1}$ for the
GREG-adjusted estimator is based on weight trimming. Whereas the GREG
estimator of \citet{Lehtonen1998} in (\ref{eqtgr}) is unbiased,
the proposed estimator in (\ref{eqygreg}) is biased. A related approach
was also presented by \citet{Beaumont2006} in the context of
missing survey data where the inclusion probabilities (design weights)
are known. They proposed a robust-GREG estimator by truncating large
weights to reduce the effect of widely dispersed design weights. In the
simulation study in Section~5, we observed that the GREG-adjusted
estimator (\ref{eqygreg}) is more efficient than the post-stratified
mean in all scenarios and more efficient than estimator~(\ref
{eqestimator}) in some scenarios. \citet{Beaumont2006} also show
that their estimator is more efficient than the standard GREG estimator
when widely dispersed design probabilities are present.\looseness=1

In the simulation study in Section~\ref{sec5} and the data example in
Section~\ref{sec6}, we used $w_0=3$ for the GREG-adjusted estimates. To
investigate the impact of this cutoff value, we performed some extra
simulations and investigated the GIS with other values of $w_0$. These
results can be found in Appendix E of the Supplementary Material [\citet{supp}]. As
expected, smaller values of $w_0$ yield estimates that have a larger
bias, but smaller variance; the opposite is true for larger values of
$w_0$. Developing systematic methods to choose the value of $w_0$ for
the GREG-adjusted estimates is an important future research endeavor.
To this purpose, existing approaches to estimate the optimal cutoff
value can be of help [see, e.g., \citet{Cox1981,Potter1990,Little1993}].

Simulation results (see Appendix~E of the Supplementary Material)
suggest that for smaller sample size ($n=150$) the estimators \textit
{npar} and \textit{npar-greg} still perform adequately but do not yield
superior MSE results. For smaller sample sizes it is much harder to fit
a nonparametric regression model to the data and this results in an
increase in the variability of the \textit{npar} and \textit{npar-greg}
estimators. Therefore, care needs to be taken when using the proposed
methods with small sample sizes.

Variance estimates of the proposed methods were described. The
analytical formula to calculate the variance ignores the uncertainty in
estimating the variance parameters. The bootstrap procedure for the
weight smoothed estimates and the jackknife resampling approach for the
GREG-adjusted\vadjust{\goodbreak} estimates were proposed as alternatives. Another approach
is turning to Bayesian methods. A $t$-based correction to account for the
uncertainty can also be considered in the analytical case [\citet
{Lazzeroni1998}]. In the simulation study, we observed that a normal
reference distribution yielded confidence intervals with appropriate coverages.

Computationally, weight smoothed estimates are more complex than the
post-stratified, unweighted and trimmed means. This complexity has two
main sources: (i)~Weight smoothed estimates are specific to each
separate survey outcome. For the analysis of another response from the
same survey, new computations are necessary. On the other hand, the
unweighted, post-stratified and trimmed mean have fixed weights for
each response in the same survey. As stated by \citet
{Lazzeroni1998}, any procedure directed at reducing variance must
tailor the weights, depending on the degree of association of the
post-stratifier with the outcome. (ii)~To calculate a weight
smoothed estimate, a GLMM must be fit which is computationally hard. A
computer intensive resampling procedure must be used to obtain variance
estimates. However, modern computing power has made these computations
practicable.

This article focused on binary survey outcomes. Nonetheless, because
the GLMM framework is used to estimate the models, extensions to other
distributions in the exponential family of distributions can be
performed without much extra work. The formulae and methods proposed in
our paper are still valid for outcomes with nonbinary distributions.
Only one post-stratifying variable was considered in this article.
Extending the weight smoothing models to more than one post-stratifier
is possible without much effort and is a topic of further research.

Model~(\ref{eqgeneralmodeltime}) used for the trend estimation is
possibly not flexible enough for the application example. There is
evidence in the influenza literature that the timings and lengths of
influenza-like illness peaks vary by age groups [see, e.g., \citet
{Adler2014}]. Model~(\ref{eqgeneralmodeltime}) assumes additive effects
of time and age groups and would thus not allow for this possible
interaction between age and time. The inclusion of such an interaction
term falls out of the scope of this paper. Nevertheless, in general,
model~(\ref{eqgeneralmodeltime}) can be extended to allow for
interaction effects.

From the results in our simulation study, we recommend the use of the
GREG-adjusted weight smoothed estimate with a NPAR model assumption. It
performed the most consistent over all simulation conditions. This
robustness comes, however, with a cost of efficiency when the true
underlying model is less complex.

\section*{Acknowledgments}
For the simulation study we used the infrastructure of the VSC ---
Flemish Supercomputer Center, funded by the Hercules\vadjust{\goodbreak} Foundation and the
Flemish Government --- Department EWI. The authors gratefully acknowledge
C.~E. Koppeschaar, R. Smallenburg, S.~P. van Noort, and the entire Dutch
and Belgian Influenzanet team for providing us with all necessary data
from the GIS in Belgium.


\begin{supplement}[id=suppA]
\stitle{Additional details and results}
\slink[doi]{10.1214/15-AOAS874SUPP} 
\sdatatype{.pdf}
\sfilename{aoas874\_supp.pdf}
\sdescription{The reader is referred to the online Supplementary
Material for more information on how the models can be cast in the
GLMM framework (Appendix A), for more details on the estimation method
(Appendix B), for annotated SAS and R programs (Appendix C), for
additional simulation results (Appendix D), and for additional results
for different values of $w_0$, additional results for smaller sample
size and results on model fits and other spline basis functions
(Appendix E).}
\end{supplement}

%

\printaddresses
\end{document}